\documentstyle[prc,aps,preprint]{revtex}
\begin{document}
\draft
\date{\today}
\title{The decay $\rho^{0}\rightarrow \pi^{0}\pi^{0}\gamma$ and
the role of $\sigma$-meson}

\author{A. Gokalp~\thanks{agokalp@metu.edu.tr} and
        O. Yilmaz~\thanks{oyilmaz@metu.edu.tr}}
\address{ {\it Physics Department, Middle East Technical University,
06531 Ankara, Turkey}}
\maketitle

\begin{abstract}
We study the radiative decay $\rho^{0}\rightarrow\pi^{0}\pi^{0}\gamma$
within the framework of a phenomenological approach in which the
contributions of $\sigma$-meson and $\omega$-meson intermediate
states and of the pion-loops are considered. We conclude that the
$\sigma$-meson amplitude provides the dominant contribution.
\end{abstract}

\thispagestyle{empty}
~~~~\\
\newpage
\setcounter{page}{1}

The radiative decays of neutral vector mesons into a single photon
and a pair of neutral pseudoscalar mesons have been a subject of
continuous interest theoretically and are becoming an area of
active experimental research with the advent of high-luminosity,
low energy $e^+e^-$-machines such as Novosibirsk and Frascati that
will make possible studying these rare decays with very small
branching ratios. Such measurements may serve as tests for the
theoretical ideas about the interesting mechanisms of these decays
because of the absence of charged particles hence the bremstrahlung
radiation, as well as sheding light on the structure of the intermediate
states involved in these decays.

The decay $\rho^{0}\rightarrow\pi^{0}\pi^{0}\gamma$ among other
radiative $\rho$-meson decays was first studied by Singer \cite{R1} assuming
that this decay proceeds through an ($w\pi$) intermediate state as
$\rho^{0}\rightarrow (w)+\pi^{0}\rightarrow(\pi^{0}\gamma)+\pi^{0}$.
Renard \cite{R2} later considered $\sigma$-meson intermediate state
in addition to this intermediate vector meson dominance (VDM)
mechanism and noted the dependence of the decay rate as well as
the photon spectrum on $\sigma$-meson mass.

The radiative decay processes of the type
$V^{0}\rightarrow P^{0}P^{0}\gamma$ where P and V belong to the
lowest multiplets of vector (V) and pseudoscalar (P) mesons were
studied by Fajfer and Oakes \cite{R3} using a low energy effective
Lagrangian approach with gauged Wess-Zumino terms. They neglected
scalar meson contributions and obtained the branching ratio as
B($\rho^{0}\rightarrow\pi^{0}\pi^{0}\gamma$)=$2.89\times 10^{-5}$.
The contribution of intermediate vector mesons to the decays
$V^{0}\rightarrow P^{0}P^{0}\gamma$ was later considered by
Bramon et al. \cite{R4} using standard Lagrangians obeying the
SU(3)-symmetry. Their results were in agreement with those of
Renard \cite{R2} when they were corrected using the then present-day
data, however they disagree with the numerical predictions of
Fajfer and Oakes \cite{R3} even if the initial expressions for
the Lagrangians were the same. Their results for the decay rate
and the branching ratio were
$\Gamma(\rho^{0}\rightarrow\pi^{0}\pi^{0}\gamma$)=$1.6$ KeV and
B($\rho^{0}\rightarrow\pi^{0}\pi^{0}\gamma$)=$1.1\times 10^{-5}$, respectively.
They also noted that final-state interactions could lead to a larger
value for the decay rate
$\Gamma (\rho^{0}\rightarrow\pi^{0}\pi^{0}\gamma$) through the chain
$\rho^{0}\rightarrow\pi^{+}\pi^{-}\gamma\rightarrow\pi^{0}\pi^{0}\gamma$.
Bramon et al. \cite{R5} lated considered these decays within the framework
of chiral effective Lagrangians and using chiral perturbation theory they
calculated the decay rates for various decays of the type
$V^{0}\rightarrow P^{0}P^{0}\gamma$ at the one-loop level, including both
$\pi\pi$ and $K\bar{K}$ intermediate loops. They showed that the one-loop
contributions are finite and to this order no counterterms are required.
In this approach, the decay
$\rho^{0}\rightarrow\pi^{0}\pi^{0}\gamma$ proceeds mainly through charged-pion
loops and Bramon et al. \cite{R5} obtained the contribution of charged-pion
loops to the decay rate as
$\Gamma(\rho^{0}\rightarrow\pi^{0}\pi^{0}\gamma$)=$1.42$ KeV
which was of the same order of magnitude as the VDM contribution calculated
by Bramon et al. \cite{R4} earlier. They also noted that the charged-kaon
loops gave a contribution to this decay rate which was $10^{3}$ times smaller
than that of charged-pion loops. They obtained the global decay rate and
the branching ratio by considering the VDM amplitude and the charged-pion
loop amplitude as
$\Gamma(\rho^{0}\rightarrow\pi^{0}\pi^{0}\gamma$)=$3.88$ KeV and
B$(\rho^{0}\rightarrow\pi^{0}\pi^{0}\gamma)$=$26\times 10^{-6}$,
respectively, and moreover
noted that the photon spectrum peaked at high photon energy $E_{\gamma}$.

Recently, radiative decays of $\rho^{0}$ and $\phi$ mesons were investigated by
Marco et al. \cite{R6} using the techniques of chiral unitary theory
developed earlier by Oller et al. \cite{R7}.
They noted that the energies of the two meson systems in the final states
of these decays are too large for these decays to be treated with standard
chiral perturbation theory, and using a chiral unitary approach they included
the final state interaction of two mesons by summing the pion-loops through
the Bethe-Salpeter equation. They obtained the branching ratio for the decay
$\rho^{0}\rightarrow\pi^{0}\pi^{0}\gamma$ as
B$(\rho^{0}\rightarrow\pi^{0}\pi^{0}\gamma$)=$1.4\times 10^{-5}$, and noted
that this result could be interpreted as
$\rho^{0}\rightarrow\gamma\sigma (\pi^{0}\pi^{0})$  since  $\pi^0\pi^0$
interaction is dominated by the $\sigma$-pole in the relevant energy regime
of this decay. Furthermore, they also pointed out that if the product of the
coupling constants F$_{V}$ and G$_{V}$ is chosen as negative $F_{V}G_{V}<0$,
where the coupling constant G$_{V}$ is deduced from
$\rho^{0}\rightarrow\pi^{+}\pi^{-}$ decay and the coupling constant
F$_{V}$ from $\rho^{0}\rightarrow e^{+}e^{-}$, then the result for
the branching ratio becomes
B$(\rho^{0}\rightarrow\pi^{0}\pi^{0}\gamma$)=$1.0\times 10^{-4}$;
therefore the measurement of this branching ratio may serve as a test
for the sign of the $F_{V}G_{V}$ product.

In this work, we follow a phenomenological approach and attempt to
calculate the decay rate for the decay
$\rho^{0}\rightarrow\pi^{0}\pi^{0}\gamma$ by considering the
$\sigma$-pole and $w$-pole diagrams as well as the charged-pion loops.
We use the coupling constant g$_{\rho\sigma\gamma}$ that we estimated
\cite{R8} using the experimental value of the branching ratio for the
radiative decay $\rho^{0}\rightarrow\pi^{+}\pi^{-}\gamma$.
Since the coupling constant g$_{\rho\sigma\gamma}$ was estimated
in terms of the $\sigma$-meson parameters M$_{\sigma}$ and $\Gamma_{\sigma}$,
we obtain the decay rate
$\Gamma(\rho^{0}\rightarrow\pi^{0}\pi^{0}\gamma)$ as a function of
these parameters and for both signs of g$_{\rho\sigma\gamma}$.

Our calculation is based on the Feynman diagrams shown in Fig. 1.
We show the amplitude for $\sigma$-meson intermediate state in Fig. 1(a).
We describe the $\rho\sigma\gamma$-vertex by the effective Lagrangian
\cite{R9}
\begin{eqnarray}
{\cal L}^{int.}_{\rho\sigma\gamma}=\frac{e}{M_{\rho}}g_{\rho\sigma\gamma}
   [\partial^{\alpha}\rho^{\beta}\partial_{\alpha}A_{\beta}
   -\partial^{\alpha}\rho^{\beta}\partial_{\beta}A_{\alpha}]\sigma~~,
\end{eqnarray}
which also defines the coupling constant $g_{\rho\sigma\gamma}$. For the
$\sigma\pi\pi$-vertex we use the effective Lagrangian \cite{R10}
\begin{eqnarray}
{\cal L}^{int}_{\sigma\pi\pi}=
\frac{1}{2}g_{\sigma\pi\pi}M_{\sigma}\vec{\pi}\cdot\vec{\pi}\sigma~~.
\end{eqnarray}
The decay width of the $\sigma$-meson that follows from this effective
Lagrangian is given as
\begin{eqnarray}
\Gamma_{\sigma}\equiv\Gamma(\sigma\rightarrow\pi\pi)=
\frac{g^{2}_{\sigma\pi\pi}}{4\pi}\frac{3M_{\sigma}}{8}
\left [ 1-(\frac{2M_{\pi}}{M_{\sigma}})^{2}\right ] ^{1/2}~~.
\end{eqnarray}
In our calculation of the invariant amplitude, in the $\sigma$-meson
propagator in Fig. 1(a) we make the replacement
$M_{\sigma}\rightarrow M_{\sigma}-\frac{1}{2}i\Gamma_{\sigma}$,
where $\Gamma_{\sigma}$ is given by Eq. (3). Since the experimental
candidate for $\sigma$-meson $f_{0}$(400-1200) has
a width (600-1000) MeV \cite{R11},
in our previous work \cite{R8}, using the experimental value of
the branching ratio for the radiative decay
$\rho^{0}\rightarrow\pi^{+}\pi^{-}\gamma$ we estimated
the coupling constant g$_{\rho\sigma\gamma}$ for a set of values of
$\sigma$-meson parameters M$_{\sigma}$ and $\Gamma_{\sigma}$.
In our present work, we use these values of the coupling constant
g$_{\rho\sigma\gamma}$. The $\omega\rho\pi$-vertex in Fig. 1(b) is
described by the effective Lagrangian \cite{R12}
\begin{eqnarray}
{\cal L}^{int.}_{\omega\rho\pi}=\frac{g_{\omega\rho\pi}}{M_{\omega}}
\epsilon^{\mu\nu\alpha\beta}\partial_{\mu}\omega_{\nu}
\partial_{\alpha}\rho_{\beta}\pi~~.
\end{eqnarray}
The coupling constant $g_{\omega\rho\pi}$ can then be determined
using Vector Meson Dominance and current-field identities \cite{R12} as
$g_{\omega\rho\pi}^{2}\simeq 84$. The $\omega\pi\gamma$-vertex
in Fig. 1(b) is described by the effective Lagrangian \cite{R12}
\begin{eqnarray}
{\cal L}^{int.}_{\omega\pi\gamma}=\frac{e}{M_{\omega}}g_{\omega\pi\gamma}
\epsilon^{\mu\nu\alpha\beta}\partial_{\mu}\omega_{\nu}
\partial_{\alpha}A_{\beta}\pi~~.
\end{eqnarray}
The coupling constant $g_{\omega\pi\gamma}$ is then obtained from the
experimental partial width \cite{R12} of the radiative decay
$\omega\rightarrow\pi^{0}\gamma$ as $g_{\omega\pi\gamma}^{2}=3.315$.
The $\rho\pi\pi$-vertex in Fig. 1(c-e) is described by the effective
Lagrangian \cite{R10}
\begin{eqnarray}
{\cal L}^{int.}_{\rho\pi\pi}=g_{\rho\pi\pi}
\vec{\rho}_{\mu}\cdot(\partial^{\mu}\vec{\pi}\times\vec{\pi})~~.
\end{eqnarray}
The experimental decay width \cite{R11} then yields the value
$\frac{g^{2}_{\rho\pi\pi}}{4\pi}=2.91$ for the coupling constant
$g_{\rho\pi\pi}$.
We describe the $\pi^{4}$-vertex in Fig. 1(c-e) by the effective
Lagrangian
\begin{eqnarray}
{\cal L}^{int.}=\frac{\lambda}{4}~(\vec{\pi}\cdot\vec{\pi})^{2}~~.
\end{eqnarray}
We note that this effective interaction results in only isospin
I=0 amplitudes. The small I=2 amplitudes were also neglected in previous
calculations \cite{R6} within the framework of chiral Lagrangians
utilizing chiral unitary theory.
We choose the coupling constant $\lambda$ as $\lambda=-
\frac{g_{\pi NN}^2}{4}\frac{M_{\sigma}^2-M_{\pi}^{2}}{M_{N}^2}$
as it is obtained in the $\sigma$-model with spontaneous symmetry
breaking \cite{R13} and we use $\frac{g_{\pi NN}^{2}}{4\pi}=14$.
We note that our effective Lagrangians
${\cal L}^{int.}_{\sigma\pi\pi}$ and ${\cal L}^{int.}_{\rho\pi\pi}$
are the ones that result from an extention of the $\sigma$ model to
include the isovector $\rho$ through a Yang-Mills local gauge theory
based on isospin with the vector meson mass generated through the
Higgs mechanism \cite{R14}. Finally, we note that the last of the
loop diagrams in Fig. 1(c-e) results from minimal coupling and
ensures gauge invariance of the pion-loop amplitudes. Similar loop
integrals were evaluated by Lucio and Pestiau \cite{R15} using
dimensional regularization and their calculations were confirmed
by Close et al. \cite{R16}. Oller \cite{R17} used their result
in his calculation of the rate of the radiative decay
$\phi\rightarrow K^{0}\bar{K}^{0}\gamma$ through an intermediate
$K^{+}K^{-}$-loop. In our case the contribution of the pion-loop
amplitudes corresponding to diagrams in Fig. 1(c-e) becomes
\begin{eqnarray}
M_{\pi}=-\frac{e~g_{\rho\pi\pi}~\lambda}{2\pi^{2}M_{\pi}^{2}}~I(a,b)
         \left [ (p\cdot k)(\epsilon\cdot u)-(p\cdot\epsilon)(k\cdot u)\right ]
\end{eqnarray}
with $a=M_{\rho}^{2}/M_{\pi}^{2}$, $b=(p-k)^{2}/M_{\pi}^{2}$,
p(u) and k($\epsilon$) being the momentum (polarization vector) of
$\rho$-meson and photon, respectively, and
\begin{eqnarray}
I(a,b)=\frac{1}{2(a-b)}
-\frac{2}{(a-b)^{2}}\left [ f(\frac{1}{b})-f(\frac{1}{a})\right ]
+\frac{a}{(a-b)^{2}}\left [ g(\frac{1}{b})-g(\frac{1}{a})\right ]
\end{eqnarray}
where
\begin{eqnarray}
&&f(x)=\left \{ \begin{array}{rr}
           -\left [ \arcsin (\frac{1}{2\sqrt{x}})\right ]^{2}~,& ~~x>\frac{1}{4} \\
\frac{1}{4}\left [ \ln (\frac{\eta_{+}}{\eta_{-}})-i\pi\right ]^{2}~, & ~~x<\frac{1}{4}
            \end{array} \right.
\nonumber \\
&& \nonumber \\
&&g(x)=\left \{ \begin{array}{rr}
        (4x-1)^{\frac{1}{2}} \arcsin(\frac{1}{2\sqrt{x}})~, & ~~ x>\frac{1}{4} \\
 \frac{1}{2}(1-4x)^{\frac{1}{2}}\left [\ln (\frac{\eta_{+}}{\eta_{-}})-i\pi \right ]~, & ~~ x<\frac{1}{4}
            \end{array} \right. 
\nonumber \\
&& \nonumber \\
&&\eta_{\pm}=\frac{1}{2x}\left [ 1\pm(1-4x)^{\frac{1}{2}}\right ] ~.
\end{eqnarray}

In terms of the invariant amplitude ${\cal M}$(E$_{\gamma}$, E$_{1}$),
the differential decay probability  for
$\rho^{0}\rightarrow\pi^{0}\pi^{0}\gamma$ for an unpolarized
$\rho^{0}$-meson at rest is then given as
\begin{eqnarray}
\frac{d\Gamma}{dE_{\gamma}dE_{1}}=\frac{1}{(2\pi)^{3}}~\frac{1}{8M_{\rho}}~
\mid {\cal M}\mid^{2} ,
\end{eqnarray}
where E$_{\gamma}$ and E$_{1}$ are the photon and pion energies respectively.
We perform an average over the spin states of
$\rho^{0}$-meson and a sum over the polarization states of the photon.
The decay width $\Gamma(\rho\rightarrow\pi^{0}\pi^{0}\gamma)$ is then
obtained by integration
\begin{eqnarray}
\Gamma=\frac{1}{2}\int_{E_{\gamma,min.}}^{E_{\gamma,max.}}dE_{\gamma}
       \int_{E_{1,min.}}^{E_{1,max.}}dE_{1}\frac{d\Gamma}{dE_{\gamma}dE_{1}}
\end{eqnarray}
where now the factor ($\frac{1}{2}$) is included because of the
$\pi^{0}\pi^{0}$ in the final state.
The minimum photon energy is E$_{\gamma, min.}=0$ and the maximum photon
energy is given as
$E_{\gamma,max.}=(M_{\rho}^{2}-4M_{\pi}^{2})/2M_{\rho}$=338 MeV.
The maximum and minimum values for pion energy E$_{1}$ are given by
\begin{eqnarray}
\frac{1}{2(2E_{\gamma}M_{\rho}-M_{\rho}^{2})}
[ -2E_{\gamma}^{2}M_{\rho}+3E_{\gamma}M_{\rho}^{2}-M_{\rho}^{3}
 ~~~~~~~~~~~~~~~~~~~~~~~~~~~~ \nonumber \\
\pm  E_{\gamma}\sqrt{(-2E_{\gamma}M_{\rho}+M_{\rho}^{2})
       (-2E_{\gamma}M_{\rho}+M_{\rho}^{2}-4M_{\pi}^{2})}~] ~.
\nonumber
\end{eqnarray}

We present the results of our calculation in Table 1.
Since we use the coupling constant $g_{\rho\sigma\gamma}$ that
we estimated using the experimental decay rate for the radiative
decay $\rho^{0}\rightarrow \pi^{+}\pi^{-}\gamma$
for a set of values of $\sigma$-meson parameters M$_{\sigma}$ and
$\Gamma_{\sigma}$ in our previous work \cite{R8}, we also obtain
the decay rate $\Gamma(\rho^{0}\rightarrow \pi^{0}\pi^{0}\gamma$)
as a function of these parameters.
We also consider for $\sigma$ meson parameters the values suggested by
two recent experiments, that is
M$_{\sigma}=555$ MeV  $\Gamma_{\sigma}=540$ MeV from CLEO \cite{R18},
and M$_{\sigma}=478$ MeV  $\Gamma_{\sigma}=324$ MeV from Fermilab E791
\cite{R19}.
Since in our previous work
we obtained a quadric equation for the coupling constant
$g_{\rho\sigma\gamma}$ which resulted in two values of this
coupling constant, one being positive and one being negative,
for a given set of $\sigma$-meson parameters M$_{\sigma}$ and
$\Gamma_{\sigma}$ we obtain two values for the decay rate
$\rho^{0}\rightarrow \pi^{0}\pi^{0}\gamma$,
one using the positive value of the coupling constant $g_{\rho\sigma\gamma}$
and one using the negative value. We therefore
note that the measurement of the decay rate
$\Gamma(\rho^{0}\rightarrow \pi^{0}\pi^{0}\gamma$)
may serve as a test for the sign of this coupling constant.
This result was also noticed by Marco et al. \cite{R6}
for the sign of the product of the coupling constants
$F_{V}G_{V}$ in their work. We further note that the measurement of
the decay rate may also help to limit the ranges of the
$\sigma$-meson parameters M$_{\sigma}$ and $\Gamma_{\sigma}$
as well.

We like to stress that within the framework of our phenomenological
approach the main contribution to the decay rate
$\Gamma(\rho^{0}\rightarrow \pi^{0}\pi^{0}\gamma$)
comes from the $\sigma$-meson intermediate state amplitude. Indeed,
for $M_{\sigma}$=478 MeV, $\Gamma_{\sigma}$=324 MeV, and
$g_{\rho\sigma\gamma}$=5.92, we obtain the value of the decay rate as
$\Gamma(\rho^{0}\rightarrow \pi^{0}\pi^{0}\gamma)=289$ KeV and
the contributions of the $\sigma$-meson intermediate state amplitude,
the $w$-meson intermediate state amplitude and the pion-loop
amplitudes are
\begin{eqnarray}
\Gamma(\rho^{0}\rightarrow \pi^{0}\pi^{0}\gamma)_{\sigma}=323~~KeV,
~~~~~~
\Gamma(\rho^{0}\rightarrow \pi^{0}\pi^{0}\gamma)_{\omega}=1.46~~KeV,
\nonumber \\
\Gamma(\rho^{0}\rightarrow \pi^{0}\pi^{0}\gamma)_{\pi}=2.7~~KeV
~~~~~~~~~~~~~~~~~~~~~~~~~~~~~
\end{eqnarray}
respectively. Our results for the $w$-meson intermediate state and pion-loop
amplitudes are in resonable agreement with the calculations of Bramon et al.
\cite{R5} and Marco et al. \cite{R6}.
The apparent discrepancy between our result for pion-loop contribution
and that of Marco et al. \cite{R6} can be resolved by noting that the
coupling constant $g_{\rho\pi\pi}$ in our work corresponds to the constant
$\frac{G_{V}M_{\rho}}{f_{\pi}^{2}}$ in their work using the framework
of chiral Lagrangians. They used the values $G_{V}=55$ MeV, $f_{\pi}=93$ MeV
resulting in the coupling constant $g_{\rho\pi\pi}$=4.9. On the other hand,
in our phenomenological approach we determine this coupling constant
from the experimental decay rate of $\rho^{0}\rightarrow\pi^{+}\pi^{-}$
\cite{R11} using the effective Lagrangian given in Eq. 6 as
$g_{\rho\pi\pi}$=6.0. If we use this coupling constant as
$g_{\rho\pi\pi}$=4.9 corresponding to
the coupling constants in the work of Marco et al. \cite{R6}, for the
contribution of the pion-loops we obtain the result
$\Gamma(\rho^{0}\rightarrow \pi^{0}\pi^{0}\gamma)_{\pi}=2.14$ KeV which
compares well with the result
$\Gamma(\rho^{0}\rightarrow \pi^{0}\pi^{0}\gamma)_{\pi}=2.11$ KeV of
Marco et al. \cite{R6}. However, since our result for
the contribution of the $\sigma$-meson amplitude is much larger than the
contributions of VDM and pion-loop amplitudes, our calculation results
in a larger value for the decay rate
$\Gamma(\rho^{0}\rightarrow \pi^{0}\pi^{0}\gamma$) than the previous
calculations which did not include the contribution coming from the
$\sigma$-meson amplitude.

Moreover, if we use the values for $\sigma$-meson parameters the
values $M_{\sigma}$=500 MeV, $\Gamma_{\sigma}$=250 MeV as used by
Soyeur \cite{R20}, and furthermore use the value of the coupling constant
$g_{\rho\sigma\gamma}$=2.71 which Titov et al. \cite{R9} obtained
from Friman and Soyeur \cite{R12} analysis of $\rho$-meson photoproduction
using VMD we obtain the decay rate as
$\Gamma(\rho^{0}\rightarrow \pi^{0}\pi^{0}\gamma)=56.2$ KeV,
or if we use $M_{\sigma}$=500 MeV, $\Gamma_{\sigma}$=500 MeV and
$g_{\rho\sigma\gamma}$=2.71 we then obtain it as
$\Gamma(\rho^{0}\rightarrow \pi^{0}\pi^{0}\gamma)=31$ KeV. In both of
these calculations we determine the coupling constant
$g_{\rho\pi\pi}$ from the experimental decay rate of
$\rho^{0}\rightarrow\pi^{+}\pi^{-}$ which in the former case results in
$g_{\rho\pi\pi}$=4.5 and in the latter case in $g_{\rho\pi\pi}$=6.4.
We like to stress, however, that in our phenomenological approach,
the results of which we present in Table 1, we try to
estimate the coupling constants we use in our calculation utilizing
the experimental information, and for the coupling constant
$g_{\rho\sigma\gamma}$ we again try to relate it to $\sigma$-meson
parameters through the decay rate
$\Gamma(\rho^{0}\rightarrow \pi^{+}\pi^{-}\gamma)$ \cite{R8}.
We also use for $\sigma$-meson parameters $M_{\sigma}$ and
$\Gamma_{\sigma}$ the experimental numbers of the $f_{0}$ (400-1200)
meson \cite{R11}, which is the candidate for $\sigma$-meson,
as summarized by T\"{o}rnqvist \cite{R21}.

The photon spectra for the decay rate of the decay
$\rho^{0}\rightarrow \pi^{0}\pi^{0}\gamma$,
which may be tested in future experiments, are plotted in Fig. 2
as a function of photon energy E$_{\gamma}$.
The parameters of the $\sigma$-meson adapted
for the numerical calculations are the same as above, that is
M$_{\sigma}$=478 MeV, $\Gamma_{\sigma}$=324 MeV,
$g_{\rho\sigma\gamma}=5.92$.
It is clearly seen that the contributions of the $\sigma$-term is
much larger than the contributions of the $w$-term, pion-loop term
as well as the interference terms. The dominant $\sigma$-term
characterizes the photon spectrum which peaks at high photon energies.
For comparision, we also plot the photon spectra for the same set
of $\sigma$-meson parameters but for the negative value of the coupling
constant $g_{\rho\sigma\gamma}=-4.49$ in Fig. 3. The general
shape of the spectrum, its peaking at high photon energy as well as
the relative contributions of different terms are the same quantitatively
as for the case with the positive value of the coupling constant
$g_{\rho\sigma\gamma}$ shown in Fig. 2.
These figures clearly show the importance of the interference terms
between $\sigma$-amplitude and pion-loop amplitude and also between
$\sigma$-amplitude and $\omega$-amplitude. These interference terms
change sign when we use negative value of the coupling constant
$g_{\rho\sigma\gamma}$ instead of its positive value and this effect
is also evident in the figures.
Therefore, the measurement of the decay rate
$\Gamma(\rho^{0}\rightarrow \pi^{0}\pi^{0}\gamma$) may clarify the nature
of the $\sigma$-meson by limiting the range of its parameters as well as
determining the sign of the coupling constant $g_{\rho\sigma\gamma}$
which plays a crutial role in analyzing photoproduction reactions
within the framework of meson-exchange mechanisms such as the
photoproduction of vector mesons on nucleons near threshold.

\begin{center}
{\bf ACKNOWLEDGMENTS}
\end{center}

We thank  M. P. Rekalo for suggesting this problem to us and for
his interest during the course of our work.


\begin{table}
\caption{The decay rate $\Gamma(\rho^{0}\rightarrow \pi^{0}\pi^{0}\gamma)$
for different $\sigma$-meson parameters}
\begin{tabular}{|c|c|c||c|c||c|c||}
$M_{\sigma}$ (MeV)
&$\Gamma_{\sigma}$ (MeV)
&$g_{\sigma\pi\pi}$
&$g_{\rho\sigma\gamma}$
&$\Gamma_{+}~(KeV)$
&$g_{\rho\sigma\gamma}$
&$\Gamma_{-}~(~KeV)$ \\
\hline
400 & 500 & 7.66 & 5.27 & 229 & -5.23 & 270  \\ \hline
450 & 500 & 6.90 & 6.19 & 256 & -5.33 & 243 \\ \hline
500 & 300 & 4.93 & 6.31 & 311 & -4.41 & 202 \\ \hline
500 & 500 & 6.36 & 7.21 & 280 & -5.49 & 224 \\ \hline
600 & 400 & 5.03 & 9.28 & 334 & -5.63 & 198 \\ \hline
555 & 540 & 6.15 & 8.59 & 294 & -6.00 & 219  \\ \hline
478 & 324 & 5.29 & 5.92 & 289 & -4.49 & 215
\end{tabular}
\end{table}

\newpage

{\bf Figure Captions:}

\begin{description}

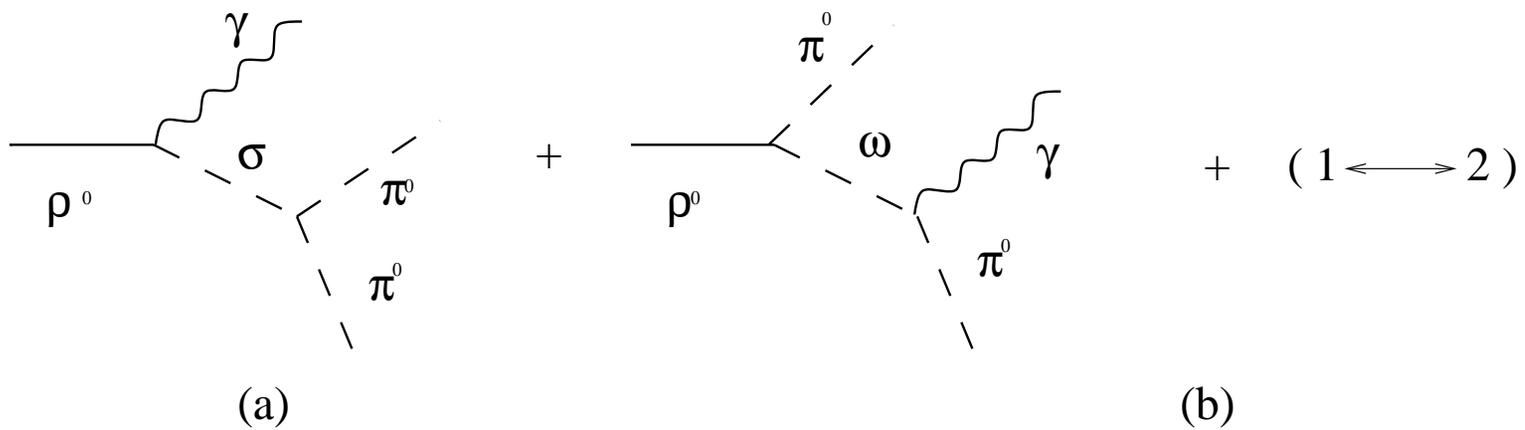
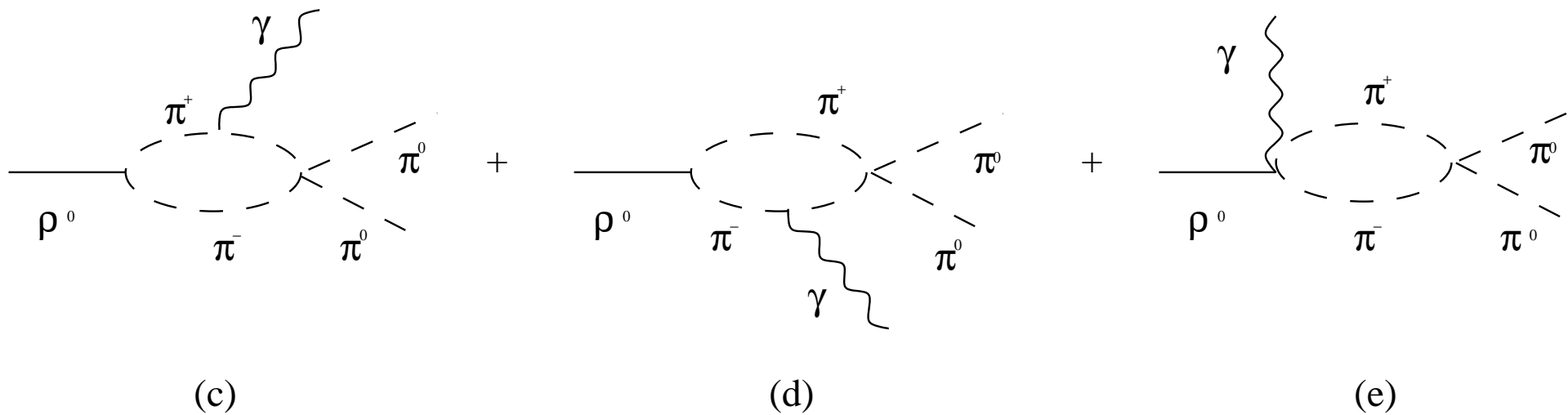
\item[{\bf Figure 1}:] Feynman Diagrams for the decay
$\rho^{0}\rightarrow \pi^{0}\pi^{0}\gamma$

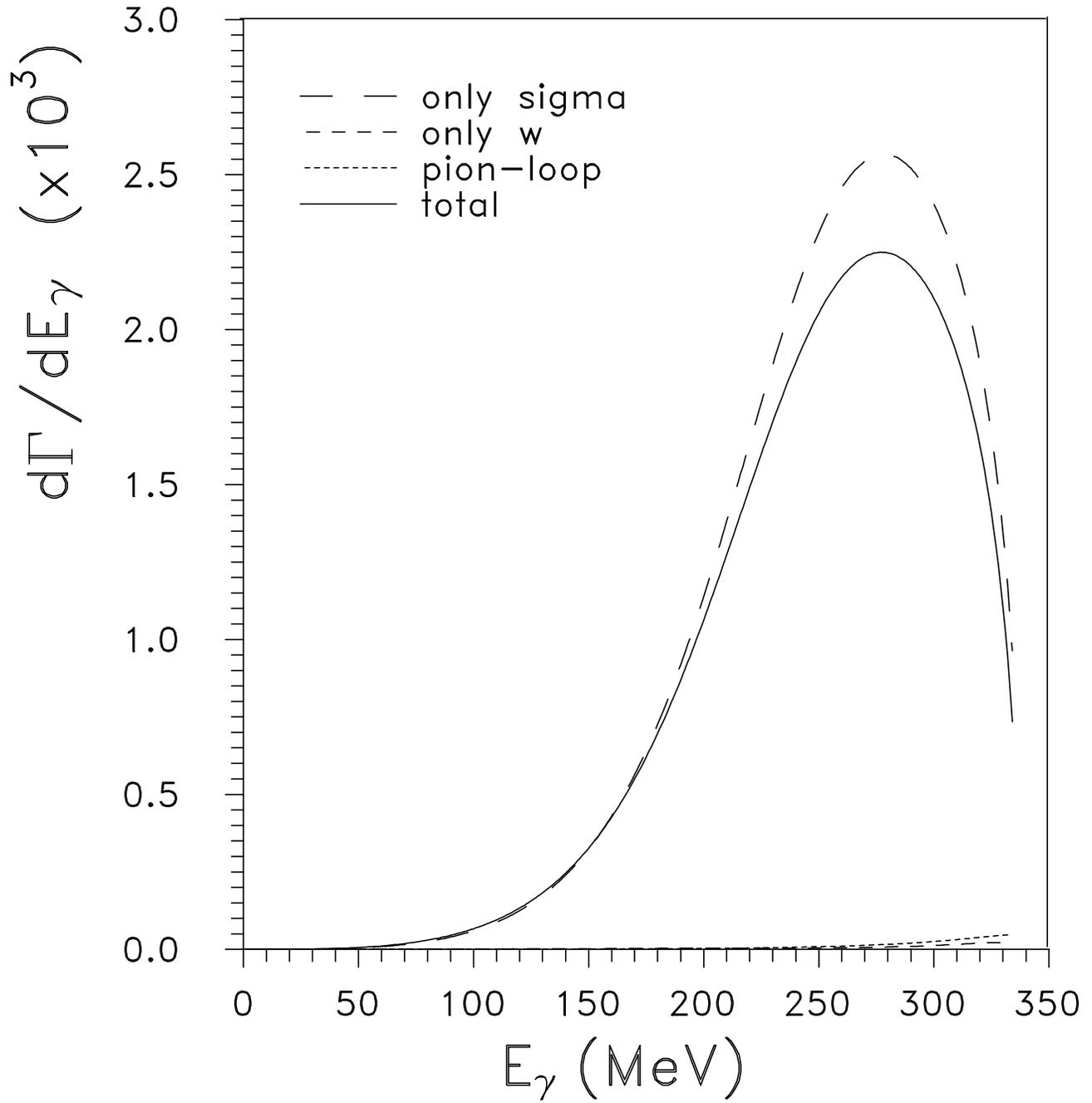
\item[{\bf Figure 2}:] The photon spectra for the decay width of
$\rho^{0}\rightarrow\pi^{0}\pi^{0}\gamma$ for $g_{\rho\sigma\gamma}>0$ .
The contributions of different terms are indicated.

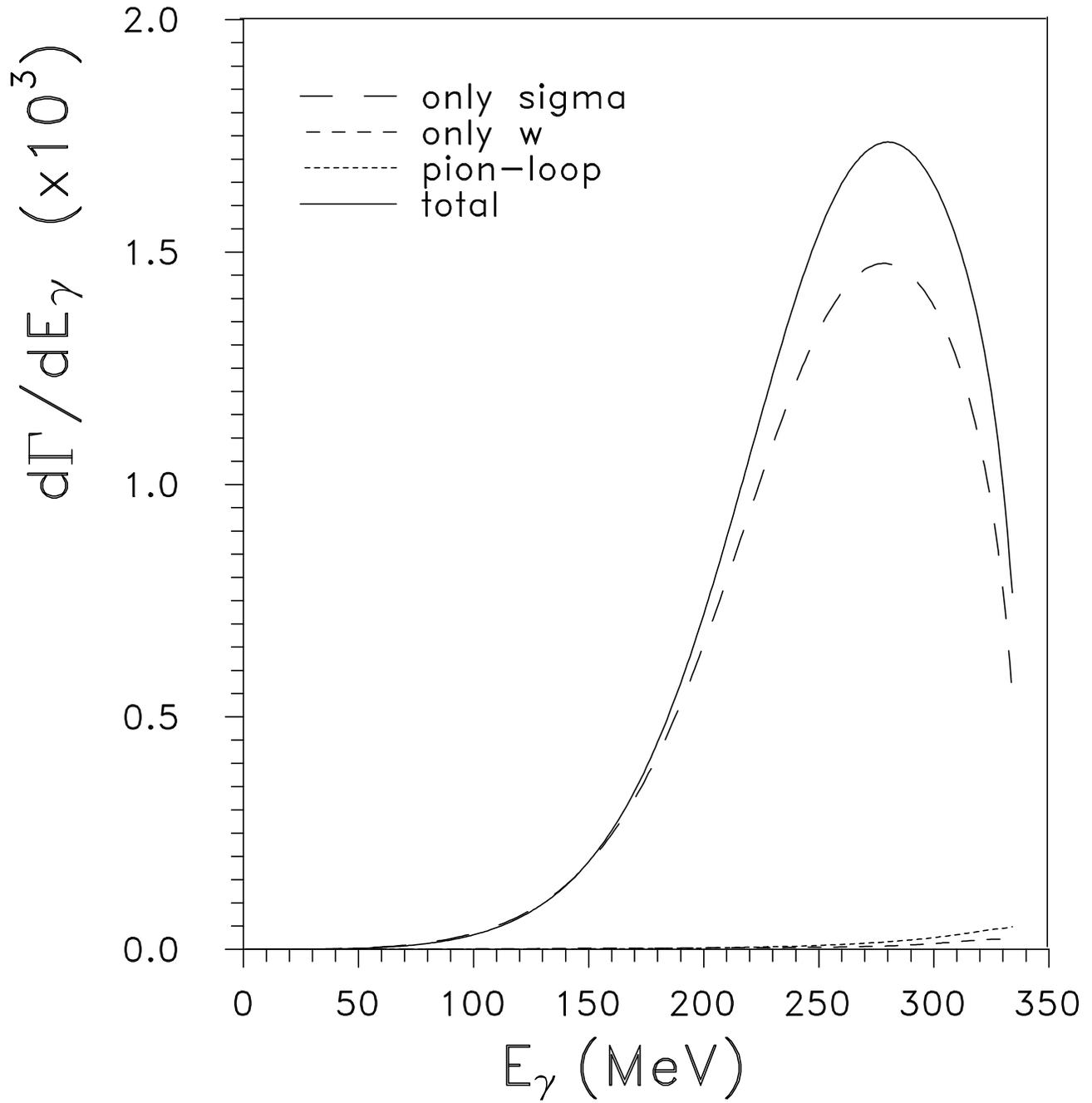
\item[{\bf Figure 3}:] The photon spectra for the decay width of
$\rho^{0}\rightarrow\pi^{0}\pi^{0}\gamma$ for $g_{\rho\sigma\gamma}<0$ .
The contributions of different terms are indicated.

\end{description}

\end{document}